\documentclass[journal]{rmaa}

\usepackage{paralist}

\usepackage{psfrag,color}

\usepackage[latin1]{inputenc}

\usepackage{amsmath}	% Advanced maths commands
\usepackage{bm}
\usepackage{ulem}

\newcommand{\mesa}{\texttt{MESA}}

\title{The effect of opacity on neutron star Type I X-ray burst quenching} 

\author{
  Mart\'in Nava-Callejas,\altaffilmark{1} 
  Yuri Cavecchi,\altaffilmark{2}
  and Dany Page\altaffilmark{1}}

\altaffiltext{1}{Instituto de Astronom\'ia, UNAM, Mexico}

\altaffiltext{2}{Universitat Polit\`ecnica de Catalunya, Spain}

\shortauthor{Nava-Callejas, Cavecchi \& Page}
\shorttitle{Opacity impact on NS burning stability}

\fulladdresses{
\item Departament de Fis\'{i}ca, EEBE, Universitat\\ Polit\`ecnica de Catalunya, Av. Eduard Maristany\\ 16, 08019 Barcelona, Spain\\ (yuri.cavecchi@upc.edu).
\item Instituto de Astronom\'ia, Universidad Nacional\\ Aut\'onoma de M\'exico, Apartado Postal\\ 70-264, 04510 M\'exico, CDMX, Mexico\\ (mnava@astro.unam.mx, page@astro.unam.mx).}

\listofauthors{M. Nava-Callejas, Y. Cavecchi, \& D. Page}
\indexauthor{Nava-Callejas, M.}
\indexauthor{Cavecchi, Y.}
\indexauthor{Page, D.}

\abstract{One long standing tension between theory and observations of Type I X-ray burst is the accretion rate at which the burst disappear due to stabilization of the nuclear burning that powers them. This is observed to happen at roughly one third of the theoretical expectations. Various solutions have been proposed, the most notable of which is the addition of a yet unknown source of heat in the upper layers of the crust, below the burning envelope. In this paper we ran several simulations using the 1D code \mesa{} to explore the impact of opacity on the threshold mass accretion rate after which the bursts disappear, finding that a higher than expected opacity in the less dense layers near the surface has a stabilizing effect.}

\resumen{Una tensi\'on largamente persistente entre teor\'ia y observaciones de estallidos de rayos-X Tipo I es la tasa de acreci\'on a la cual los estallidos desaparecen debido a la estabilizaci\'on de quemado nuclear que los origina. Se ha observado que esto ocurre aproximadamente a un tercio de las expectativas te\'oricas. Varias soluciones se han propuesto, la m\'as notable siendo la adici\'on de una fuente de calor de origen desconocido en las capas m\'as altas de la corteza, debajo de la envolvente en quemado. En este art\'iculo corrimos varias simulaciones empleando el c\'odigo 1D \mesa{} para explorar el impacto de la opacidad en la tasa de acreci\'on l\'imite arriba de la cual los estallidos desaparecen, encontrando que una opacidad m\'as alta de lo esperado en las zonas menos densas cerca de la superficie tiene un efecto estabilizador.}

\addkeyword{accretion, accretion discs}
\addkeyword{stars:neutron}
\addkeyword{X-rays: binaries}
\addkeyword{X-rays: bursts}

\begin{document}
% Typeset article header
\maketitle

%%%%%%%%%%%%%%%%%%%%%%%%%%%%%%%%%%%%%%%%%%%%%%%%%%%%%%%%%%%%%%%%%%%%%%%%%%%%%%%%%%%%%%%%%%%%%%%%%%%%%%%%%
\section{Introduction}
\label{sec:intro}
%%%%%%%%%%%%%%%%%%%%%%%%%%%%%%%%%%%%%%%%%%%%%%%%%%%%%%%%%%%%%%%%%%%%%%%%%%%%%%%%%%%%%%%%%%%%%%%%%%%%%%%%%

In binary systems hosting a neutron star and a small mass star, when the companion expands to fill its Roche lobe mass can be transferred to the compact object via an accretion disk. Under the right conditions, especially mass accretion rate, the accreted fuel on the neutron star surface will burn unstably, producing X-ray flashes known as the Type I bursts \citep[see e.g.][]{stroh_bild_2010}. The fluid begins burning in the upper layers, but the cooling proceesses are able to compensate the heating due to the nuclear reactions. As the fluid sinks deeper under the push of newer accreted layers, the burning rate increases, especially due to the increase of temperature. When the cooling cannot compensate the reaction rate anymore, the burning turns explosive and initiates the burst \citep[e.g.][]{fuji_1981,bild_1998}. Depending on the accretion rate, the first unstable ignition could be due to hydrogen or helium, and the burst can have a larger or smaller amount of hydrogen left to burn at its later stages \citep{fuji_1981,bild_1998}. Above a certain accretion rate bursts are suppressed because the fluid never reaches unstable burning conditions.

To date, one of the persistent discrepancies between theory and observations of Type I X-ray bursts is the critical accretion rate $\dot{M}_{\rm{crit}}$ above which bursts are suppressed. While observations indicate $\dot{M}_{\rm{crit}} \approx $ $0.3\dot{M}_{\rm{Edd}}$ \citep{2003A&A...405.1033C, 2007A&A...467L..33W, 2008ApJS..179..360G, 2021ASSL..461..209G}, numerical simulations employing different codes indicate $\dot{M}_{\rm{crit}}\approx 1$ to $3\dot{M}_{\rm{Edd}}$ \citep{1998ASIC..515..419B, 2007ApJ...665.1311H, Fisker_2007}, where $\dot{M}_{\rm{Edd}}$ is the Eddington accretion rate. One proposed mechanism to keep the burning layer stable is the presence of a heating source injecting up to a few MeV per baryon at densities of $10^{7}$ g cm$^{-3}$ or above, the so called \textit{shallow heating} \citep{2009ApJ...698.1020B,2017JApA...38...49W}, which is currently of unknown origin. Several explanations or alternatives have been proposed over the years, as for instance modification to the CNO break out $^{15}$O($\alpha,\gamma$)$^{19}$Ne \citep{2006ApJ...648L.123C, Davids_2011}, diffusion of $^{4}$He or rotation effects \citep{Piro2007, keek2009, 2010AstL...36..848I,cave_2020}.

In this note we show that it is possible to suppress the bursts in numerical simulations at the observed mass accretion rate by modifying the opacity. Our results indicate that this is achievable if the opacity in the layers between the surface and the ignition depth is $\gtrsim 8$ times the electron scattering opacity expected to dominate at these depth. 

%%%%%%%%%%%%%%%%%%%%%%%%%%%%%%%%%%%%%%%%%%%%%%%%%%%%%%%%%%%%%%%%%%%%%%%%%%%%%%%%%%%%%%%%%%%%%%%%%%%%
\section{Methodology}\label{sec:methods}
\label{sec:method}
%%%%%%%%%%%%%%%%%%%%%%%%%%%%%%%%%%%%%%%%%%%%%%%%%%%%%%%%%%%%%%%%%%%%%%%%%%%%%%%%%%%%%%%%%%%%%%%%%%%%

We simulated the evolution of accreted neutron star envelopes employing the public code \mesa{} v.15140 \citep{Paxton2011,Paxton_2015}. The initial profiles were constructed with an envelope code which solves the time-independent equations of stellar structure and temperature up to an inner boundary density of $\rho_{b} = 10^{9.5}$ g cm$^{-3}$ \citep{2024arXiv240313994N}.

As an inner boundary conditions for \mesa{} at $\rho_{b}$ one needs to fix the inner luminosity, $L_b$, coming from the stellar interior. Initially, to focus on the effect of opacity, we employed two different values for the base luminosity: a very low one of $L_{b} = 2.5\times 10^{-5}L_{\odot}$ and a high one of $L_{b} = 2.5\times 10^{-1}L_{\odot}$. Later, we explore the impact of changing this value by selecting a range of values up to $2.5\times 10^{3} L_{\odot}$.
It is usually assumed that this inner luminosity is controlled by the accretion rate coming from an energy release deeper in the crust with an expression of the type $L_b = Q_b \times \dot{M}/m_u$ ($m_u$ being the atomic mass unit). For the sake of comparison, our choices of $L_{b}$ would correspond, when $\dot{M} \simeq 0.3\dot{M}_{\rm{Edd}}$, to values  $Q_b = 10^{-7} - 30$ MeV baryon$^{-1}$. Following \citet{1999ApJ...524.1014S} we adopt $\dot{M}_{\rm{Edd}} = 1.1\times 10^{18}$ g s$^{-1}$ as the Eddington accretion rate.

The amount of cells of the spatial grid, as well as the time-step, are factors which might have some impact on the simulation results. In \mesa{,} the \texttt{mesh}\_\texttt{delta}\_\texttt{coeff} parameter controls the mesh refinement during a simulation: above 1.0, the number of grid cells tends to be smaller, while below 1.0 the number might reach up to 3000 cells. Unless explicitly stated, we adopt 5.0 for this coefficient.
Besides the size of the mesh, \mesa{} allows to have some control over the chosen time step. With \texttt{time}\_\texttt{delta}\_\texttt{coeff} (hereafter \texttt{tdc}), the user can ask for overall large steps in time, while \texttt{min}\_\texttt{timestep}\_\texttt{factor} (hereafter \texttt{mtf}) controls the minimum ratio between the new and the previous time step. By default, their respective values are 1.0 and 0.8. By trial and error, for some simulations we have found suitable to replace these defaults by a customized configuration of 5.0 and 1.2, which we employ unless another combination is explicitly stated.
In Appendix \ref{sec:AppA} we show that these values are a good choice and that our conclusions do not depend on it.

For the majority of the simulations we employed a customized network containing 140 species and capable of simulating an rp-process exhausting H at around $10^{6}$ g cm$^{-3}$. Dubbed \texttt{approx140}, this network ranges from $A=1$ to $A=80$ and covers H burning via CNO and rp-process, He burning via $3\alpha$ and $\alpha$-captures, C-O fusion and electroweak decays of heavy nuclides at $A>56$ as a result of the production of ashes. The list of nuclides used in \texttt{approx140} can be found in Table \ref{tab:net} and more details about it can be found in Appendix \ref{sec:AppB}. As a refractory layer at the base, we use $^{80}$Kr.

For the chemical composition of the accreted material we employed a Solar-like scheme with $70\% $ $^{1}$H, $29\%$ $^{4}$He and the remaining $1\%$ automatically distributed, among the remaining nuclides in the network, by \mesa{,} throughout the command \texttt{accretion\_zfracs} $=3$, corresponding to the distribution of metals from \citet{1998Sauval}. A little exploration on the sensibility of our results for choosing this automatized distribution or a simple mixture of $^{1}$H, $^{4}$He and $^{12}$C is explored in the Appendix.

%------------------------------------------------------------------------------------------------------------------------------------------------------------------------------------------------------------------------------------------------%
\begin{table}
	\centering
	\caption{List of nuclides in the network \texttt{approx140}.}
	\begin{tabular}{|p{0.4cm}|p{1.1cm}|p{0.5cm}|p{1.1cm}|p{0.5cm}|p{1.5cm}|}
		\hline
		$Z$ & $A$ & $Z$ & $A$ & $Z$ & $A$ \\
		\hline
		\hline
        n & 1 & S & 28-32 & Cu & 56-60\\
		H & 1 & Cl & 32-34 & Zn & 58-62,64\\
		He & 4 & Ar & 33-36 & Ga & 61-65\\
		C & 12 & K & 36-38 & Ge & 62-66,68\\
		N & 13-15 & Ca & 37-40 & As & 66-69\\
		O & 14-18 & Sc & 40-42 & Se & 68-70,72\\
		F & 17-19 & Ti & 41-44 & Br & 70-73\\
		Ne & 18-20 & V & 44-46 & Kr & 72-74,76,80\\
		Na & 20-21 & Cr & 45-48 & Rb & 74-77,80\\
		Mg & 21,22,24 & Mn & 48-50 & Sr & 76-78,80\\
		Al & 23-25 & Fe & 49-52,54 & Y & 78-80\\
		Si & 24-26,28 & Co & 52-56 & Zr & 80\\
		P & 27-30 & Ni & 53-58,60 & & \\
		\hline
	\end{tabular}\label{tab_network_net}
        \label{tab:net}
\end{table}
%------------------------------------------------------------------------------------------------------------------------------------------------------------------------------------------------------------------------------------------------%

We override the opacity using a custom routine, based on the one provided by Bill Wolf's website, \texttt{my\_other\_kap\_get}\footnote{\url{https://billwolf.space/projects/leiden_2019/}}.
The total opacity is given by $\kappa = [\kappa^{-1}_{\rm{rad}} + \kappa^{-1}_{\rm{cond}}]^{-1}$, receiving contributions from radiation, $\kappa_{\rm{rad}}$, and conduction, $\kappa_{\rm{cond}}$. For the latter, in all the simulations the routine uses the tables provided by \mesa{.} For $\kappa_{\rm{rad}}$, we considered and compared the results using values from the tables provided with the \mesa{} source code and from the extra subroutines by Wolf. These subroutines are based on the additive combination of free-free opacity from \citet{1999ApJ...524.1014S}, electron scattering from 
\citet{1983ApJ...267..315P}, and the correction factor from \citet{2001A&A...374..213P}.
We will call this opacity the fiducial opacity.
For some models we replaced Paczynski's fit with the one of \citet{2017ApJ...835..119P}, as indicated in the text.

To make sure that our conclusions are not affected by our choices for the mesh, the time step or the network, we conducted extensive tests which we report in Appendix \ref{sec:AppA}. In what follows we focus on the results based on changing the opacity.

%%%%%%%%%%%%%%%%%%%%%%%%%%%%%%%%%%%%%%%%%%%%%%%%%%%%%%%%%%%%%%%%%%%%%%%%%%%%%%%%%%%%%%%%%%%%%%%%%%%%
\section{Results}\label{sec:results}
%%%%%%%%%%%%%%%%%%%%%%%%%%%%%%%%%%%%%%%%%%%%%%%%%%%%%%%%%%%%%%%%%%%%%%%%%%%%%%%%%%%%%%%%%%%%%%%%%%%%

%%%%%%%%%%%%%%%%%%%%%%%%%%%%%%%%%%%%%%%%%%%%%%%%%%%%%%%%%%%%%%%%%%%%%%%%%%%%%%%%%%%%%%%%%%%%%%%%%%%%
\subsection{Increased/decreased opacity at fixed accretion rate}\label{subsec:first}
%%%%%%%%%%%%%%%%%%%%%%%%%%%%%%%%%%%%%%%%%%%%%%%%%%%%%%%%%%%%%%%%%%%%%%%%%%%%%%%%%%%%%%%%%%%%%%%%%%%%

%------------------------------------------------------------------------------------------------------------------------------------------------------------------------------------------------------------------------------------------------%
\begin{figure}
\centering
\includegraphics[width=0.36\textwidth]{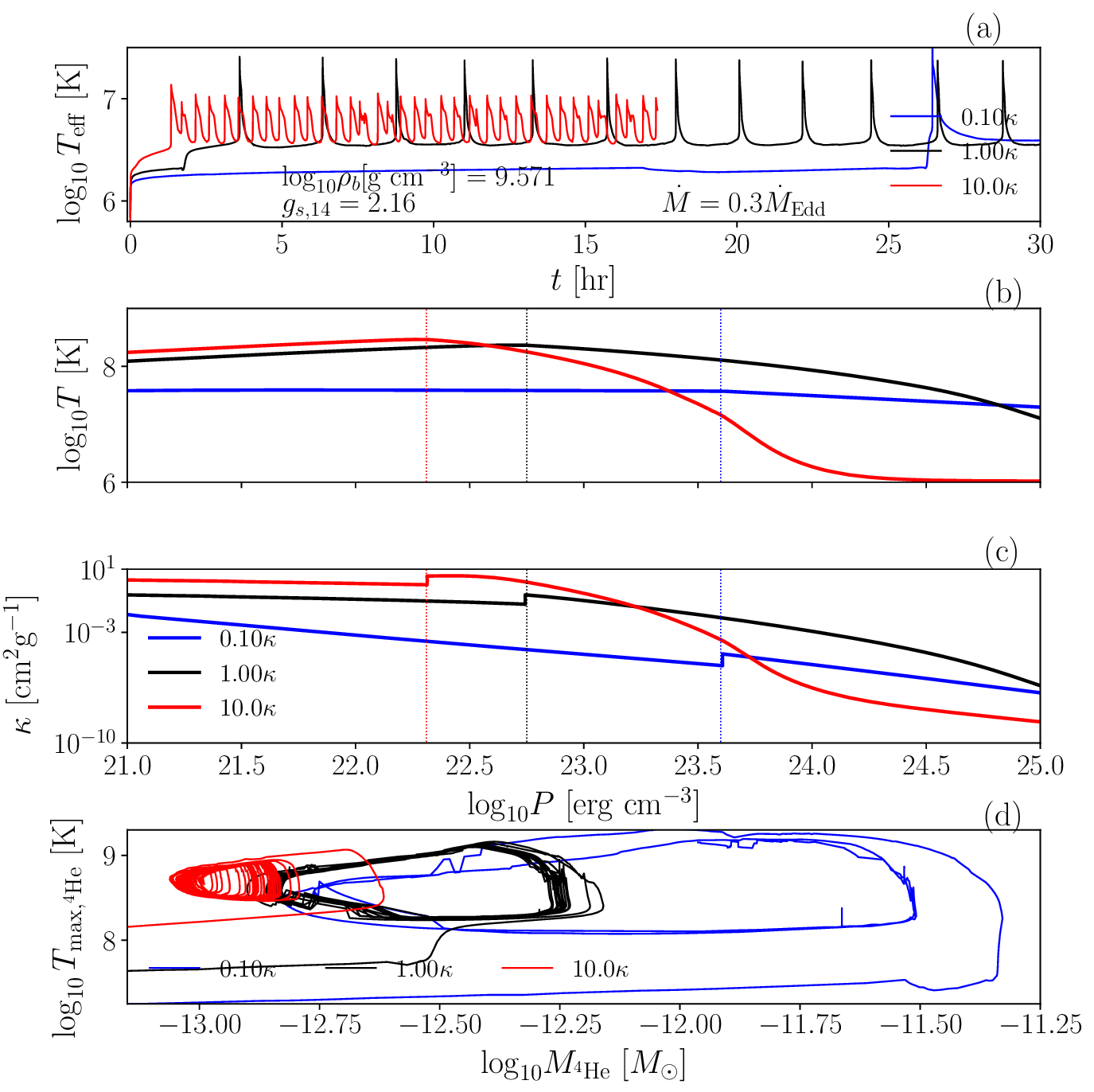}
\caption{Effects of a change in opacity $\kappa$ at a fixed mass accretion rate of $\dot{M} = 5.26\times 10^{-9} M_\odot$ yr$^{-1} = 0.3 \, \dot{M}_{\rm{Edd}}$. We consider three models with $\kappa$ unaltered or multiplied or divided by a factor of 10.
Panel (a): time evolution of $T_{\rm{eff}}$.
Panel (b): temperature profiles just before the first explosion.
Panel (c): opacity profiles at the same time.
Panel (d): temperature vs total mass evolution of the helium layer.
In panels (b) and (c) the three dotted vertical lines indicate the depth at with the explosion is occurring. The luminosity at the base for the three models is $L_{b} = 2.5\times 10^{-5} L_{\odot}$, corresponding to $Q_{b} = 3\times 10^{-7}$ MeV baryon$^{-1}$.}
\label{fig:kappa1}
\end{figure}
%------------------------------------------------------------------------------------------------------------------------------------------------------------------------------------------------------------------------------------------------%

We first analyze the impact of applying an overall opacity factor.
We consider a fixed mass accretion rate, $5.26 \times 10^{-9}\ M_{\odot}\, \rm{yr}^{-1}$ (roughly corresponding to $0.3 \dot{M}_{\rm{Edd}}$) and show the results in Fig.~\ref{fig:kappa1}.
One sees from panel (a) that the base model, with the fiducial opacity, exhibits a typical bursting behavior, with a recurrence time of the order of 2.5 hours, after an initial heating phase.
With $\kappa$ reduced by a factor of 10 it takes much longer for the first burst to appear while with the 10 times larger opacity the bursting behavior is much accelerated, similar to millihertz oscillations \citep{pacz_1983,revni_2001,2007ApJ...665.1311H}.

Panel (b), showing the temperature profiles just before explosion, explains this difference in behaviours. Increased opacity keeps the burning layer warmer, allowing it to explode at lower density. In contradistinction, the lowered opacity implies a very effective dissipation of the released nuclear energy, resulting in much lower temperatures, with little density dependence, thus forcing the accreted matter to reach higher densities before it can explode at a much delayed time.

In panel (c) dramatic changes in opacity are visible. As expected from our arbitrary 10 - 1/10 altering factor, we observe a global increase/decrease of opacity with respect to the fiducial one, although the effect is more pronounced in the 1/10 reduction scenario than in the 10-times amplified one. This is a consequence of the induced differences in nuclear burning: 
the colder 1/10 decreased opacity produces less heavy elements through the rp-process and thus sees its opacity further reduced by having more abundant low Z nuclei.
The sharp transition in opacity observed in the three models, nevertheless, still marks the transition from a low-$Z$ to a high-$Z$ region between accreted and compressed matter.

Finally, in panel (d) we display the time evolution of the total mass of accreted helium, $M_{\rm{^4He}}$, and the temperature at its maximum density, $T_{\rm{max, ^4He}}$ clearly exhibiting a cyclic behavior.
In the fiducial model, explosions are triggered when $M_{\rm{^4He}}$ reaches $\sim 10^{-12.25} M_\odot$ at temperatures $T \sim 10^{8.3}$ K and the exploding layer heats up reaching $\sim 10^{9.1}$ K: subsequently $M_{\rm{^4He}}$ rapidly decreases, helium being consumed, $T$ decreases as well and the cycle resumes (compare to \citealt{2007ApJ...665.1311H}).
In the increased opacity case, one clearly sees that the higher temperatures trigger the explosion at lower densities, the maximum temperature reached is however lower. In the lowered opacity case the contrary is happening: the cycles are pushed to much higher densities and higher temperatures are reached during the bursts.

%%%%%%%%%%%%%%%%%%%%%%%%%%%%%%%%%%%%%%%%%%%%%%%%%%%%%%%%%%%%%%%%%%%%%%%%%%%%%%%%%%%%%%%%%%%%%%%%%%%%
\subsection{Accretion rate for stabilization at 10 times the reference opacity}\label{subsec:second}
%%%%%%%%%%%%%%%%%%%%%%%%%%%%%%%%%%%%%%%%%%%%%%%%%%%%%%%%%%%%%%%%%%%%%%%%%%%%%%%%%%%%%%%%%%%%%%%%%%%%

\begin{figure}
% FIGURE 2
\includegraphics[width=0.40\textwidth]{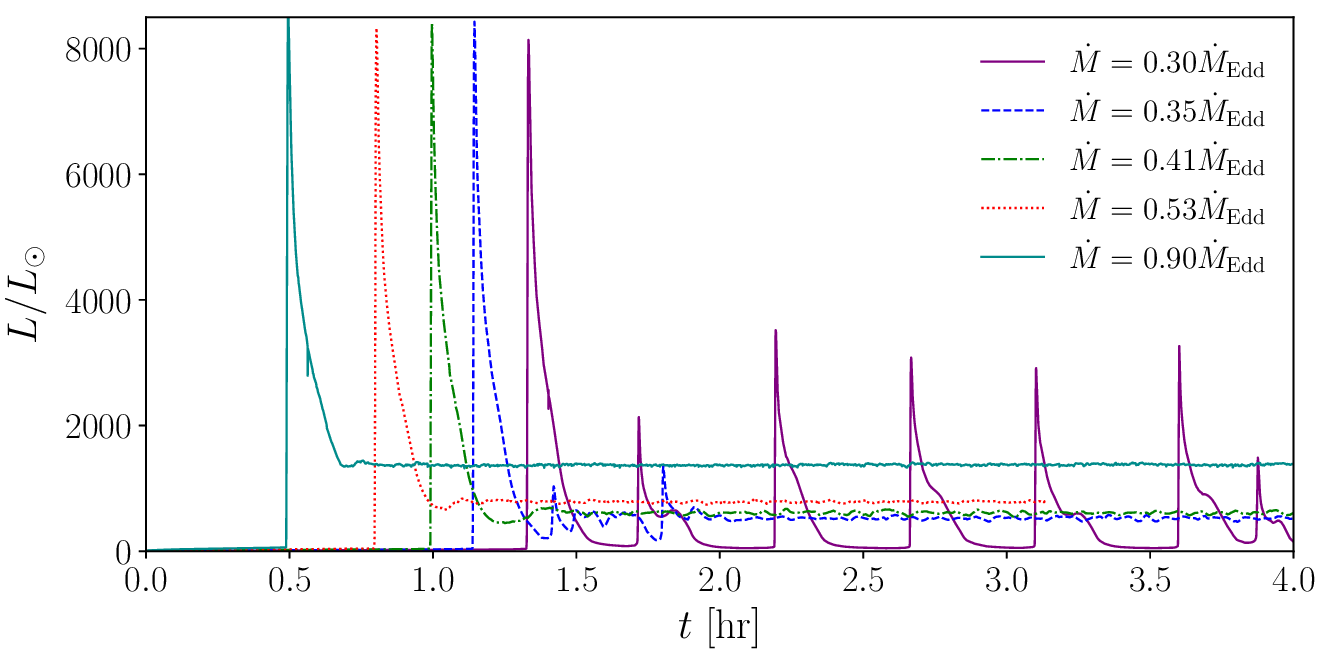}
\includegraphics[width=0.40\textwidth]{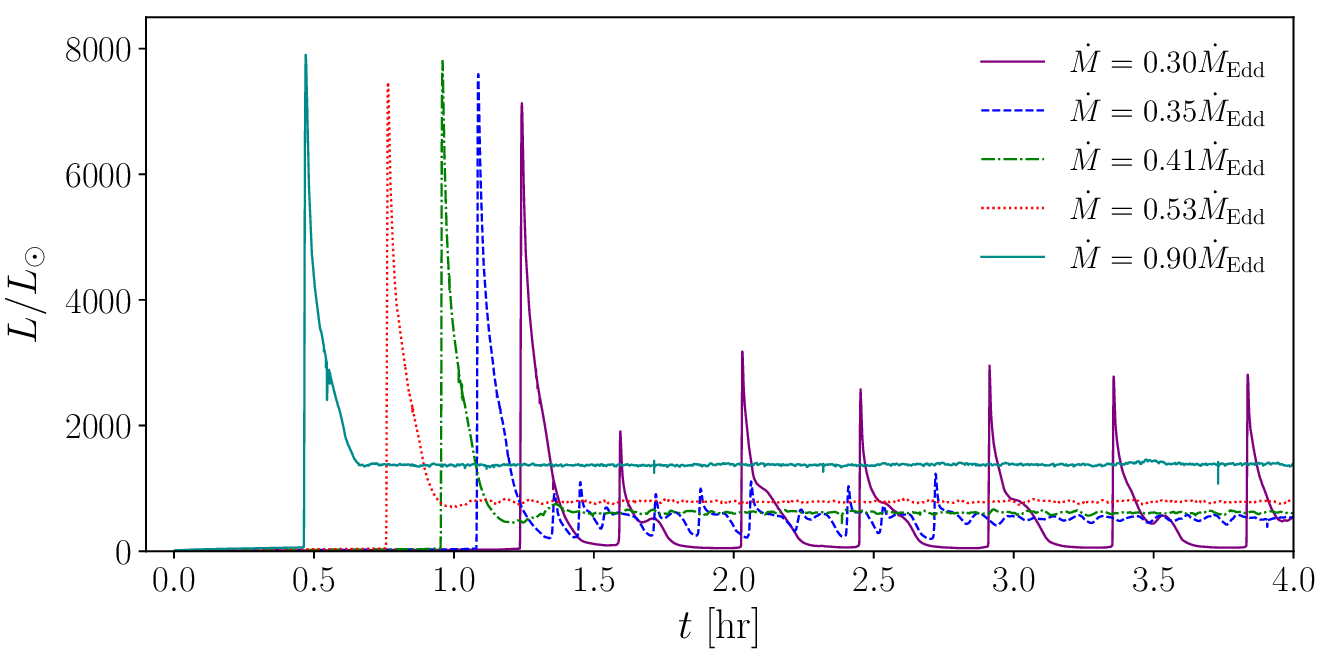}
\caption{Luminosity (in units of $L_{\odot}$) as a function of time for different values of mass accretion rate. 
The luminosity at the base is $L_{b} = 2.5\times 10^{-5} L_{\odot}$ in the upper panel
and $L_{b} = 2.5 \times 10^{-1} L_\odot$ in the lower one
(equivalent $Q_b$ ranges, corresponding to the $\dot{M}$ range exlpored, are $1-3$ $\times 10^{-7}$ and $1-3$ $\times 10^{-3}$ MeV per baryon in the upper and lower panels, respectively). In all cases the opacity is globally increased by a factor of 10.
}
 \label{fig:kappa2}
\end{figure}

Usually, in numerical simulations, the recurrence time between bursts decreases with increasing accretion rate, until the critical rate for stabilization is reached \citep[see e.g.][]{2007ApJ...665.1311H}. Considering the behavior of the $0.3 \dot{M}_{\rm{Edd}}$ envelope model at $10\kappa$, we kept this higher opacity fixed and varied the accretion rate. As shown in the upper panel of Fig.~\ref{fig:kappa2}, the bursting behavior actually ceases between 5.26 and 6.26 $\times 10^{-9}\ M_{\odot}\, \rm{yr}^{-1}$, corresponding to 0.3 and $0.35 \times \dot{M}_{\rm{Edd}}$ respectively. 
Considering the $0.35 \times \dot{M}_{\rm{Edd}}$ case, after a first burst, which should be regarded as an artifact of the simulations, the luminosity shows damped oscillations and in less than 1 hr reaches an equilibrium value. 
As intuitively expected, this equilibrium value depends on $\dot{M}$.

%------------------------------------------------------------------------------------------------------------------------------------------------------------------------------------------------------------------------------------------------%
%------------------------------------------------------------------------------------------------------------------------------------------------------------------------------------------------------------------------------------------------%

We ran more simulations between 0.5 and 1.0 $\dot{M}_{\rm{Edd}}$ displayed in the upper panel of Fig.~\ref{fig:kappa2} as well, and found that indeed all show stable burning. We also tested the effect of increasing the base luminosity to $L_{b} = 2.5 \times 10^{-1} L_\odot$. The resulting luminosities as function of time are displayed in the lower panels of Fig.~\ref{fig:kappa2}. 
Despite the small differences - at fixed accretion rate, among the models of upper and lower panels - in both scenarios the burning stabilizes between 0.3 and 0.35 $\dot{M}_{\rm{Edd}}$, as found in observations.

%%%%%%%%%%%%%%%%%%%%%%%%%%%%%%%%%%%%%%%%%%%%%%%%%%%%%%%%%%%%%%%%%%%%%%%%%%%%%%%%%%%%%%%%%%%%%%%%%%%%
\subsection{The role of the different contributions to the opacity}\label{subsec:fifth}
%%%%%%%%%%%%%%%%%%%%%%%%%%%%%%%%%%%%%%%%%%%%%%%%%%%%%%%%%%%%%%%%%%%%%%%%%%%%%%%%%%%%%%%%%%%%%%%%%%%%

\begin{figure}
\includegraphics[width=0.40\textwidth]
    {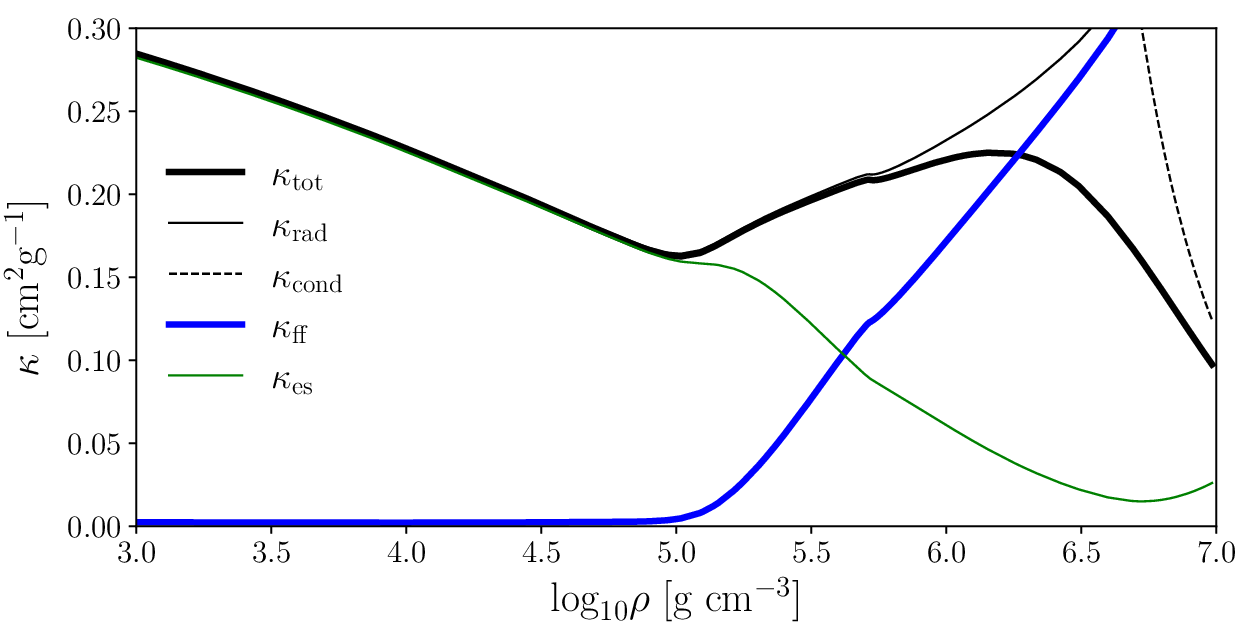}
 \caption{Opacity profile for a typical stationary accreted envelope at $\dot{M}_{\rm{Edd}}$.}
 \label{fig:kappa6}
\end{figure}

All models in the former sections have a common factor applied to the whole opacity function at all depths.
However, the opacity contains several components, from electrons and from photons, and for the latter contributions from electron scattering and free-free.
In Fig.~\ref{fig:kappa6} we exhibit a typical profile of $\kappa$ and the contributions of its various components.
Here, we will explore the impact of each component as well as the various options for the components of the radiative part.
We will designate by $\kappa_{\mathrm{Analytic}}$ the opacity where the radiative part is from the analytical fits described in Section \ref{sec:method} and by $\kappa_{\mesa{}}$ the one where we employ the \mesa{} supplied radiative opacity.
Similarly we have, in our analytical fits, two options for the electron scattering contribution which we denote as $\kappa_{\mathrm{es}}$ (Paczynski) and $\kappa_{\mathrm{es}}$ (Poutanen).

\begin{figure}
% FIGURE 3
\includegraphics[width=0.40\textwidth]
    {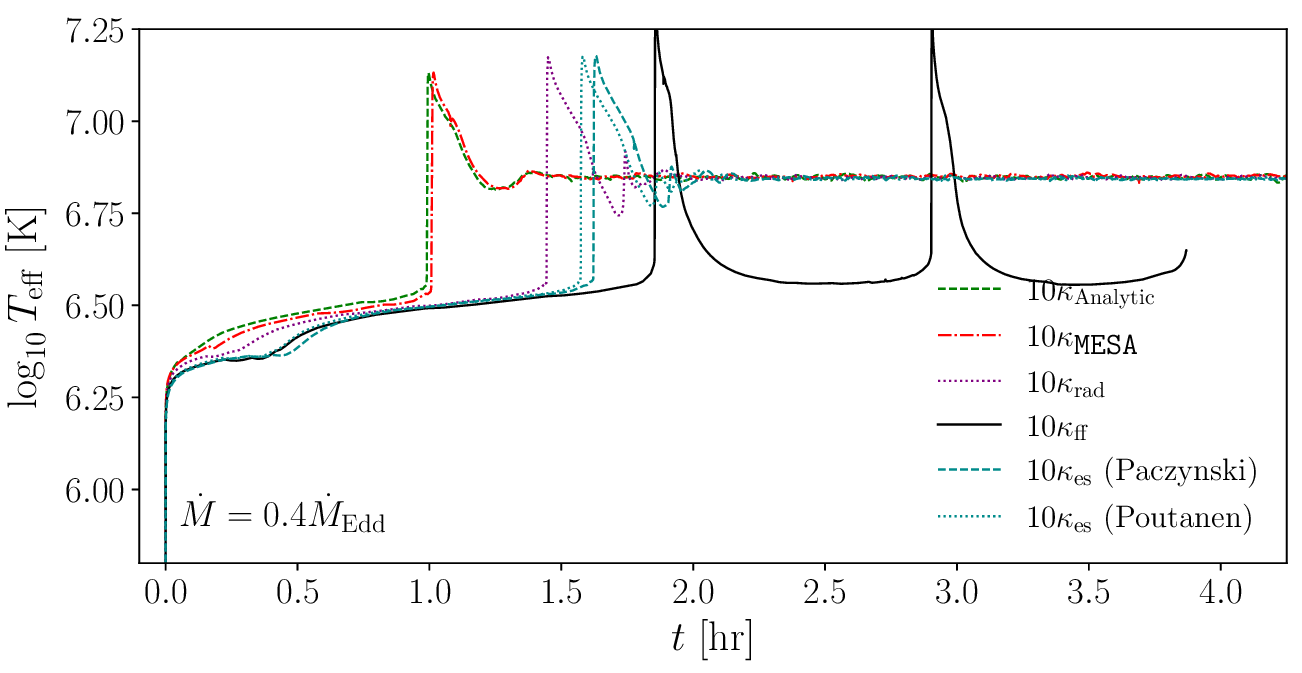}
\includegraphics[width=0.40\textwidth]
    {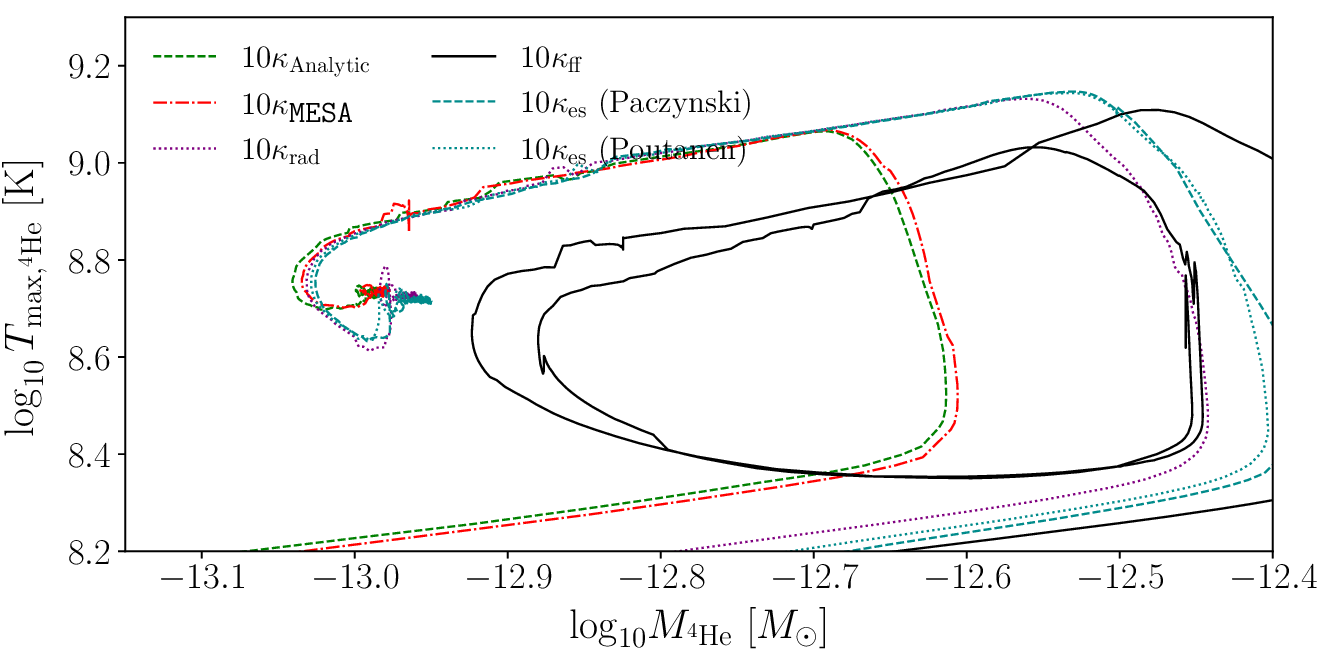}
    \caption{Upper panel: 
    effective temperature as a function of time for different opacities. 
    Lower panel:
    maximum temperature of helium matter in the envelope versus the total mass of $^{4}$He for the same models. The base luminosity is $2.5\times 10^{-5}L_{\odot}$.
    }
 \label{fig:kappa5}
\end{figure}

As a first step we compare the effect of employing the two schemes for the radiative opacity,
multiplying the whole opacity by a factor of ten.
These are the models $10 \kappa_{\mathrm{Analytic}}$ and $10\kappa_{\mesa{}}$
shown in Fig.~\ref{fig:kappa5} where one sees that the differences are minimal.

In a second step we multiply by 10 only the radiative part, leaving untouched the conductive part:
this is the model $10\kappa_{\mathrm{rad}}$ (in which we used the analytical scheme of radiative opacity).
This one also exhibits quenching of bursts with the only difference, compared to the previous models, that its initial explosion previous to quenching is delayed.
The lower panel shows how the accreted layer has to reach higher densities for the first explosion to occur: this is due to the fact that in this region the opacity is dominated by the electron opacity, which we do not alter in this case, and this allows a strong leakage of heat towards the interior, keeping these deeper layers colder than in the previous cases.
However, after this first burst, the model converges toward the same state as the previous
$10 \kappa_{\mathrm{Analytic}}$ and $10\kappa_{\mesa{}}$ runs.
This result proves that electron conduction has little effect on the bursting behavior and its possible quenching.

As a final step we determine which component of the radiative opacity is responsible for the quenching.
For this we consider the three models $10\kappa_{\mathrm{ff}}$, $10\kappa_{\mathrm{es}}$ (Paczynski), and $10\kappa_{\mathrm{es}}$ (Poutanen) in which either the free-free opacity (first model) or the electron scattering one (second and third models) is increased by a factor 10.
The last two are also exploring whether small changes in $\kappa_{\mathrm{es}}$ may have an effect
(changing from the old fit of \citet{1983ApJ...267..315P} to the recent one of \citet{2017ApJ...835..119P}).
Fig.~\ref{fig:kappa5} shows that the model $10\kappa_{\mathrm{ff}}$ does not result in stable burning while the other two do.
This definitively proves that the mechanism for the quenching of bursting behaviour is the increase in the opacity at low densities, the region where $\kappa_{\mathrm{es}}$ dominates (see Fig.~\ref{fig:kappa6}).

%------------------------------------------------------------------------------------------------------------------------------------------------------------------------------------------------------------------------------------------------%
%------------------------------------------------------------------------------------------------------------------------------------------------------------------------------------------------------------------------------------------------%

%------------------------------------------------------------------------------------------------------------------------------------------------------------------------------------------------------------------------------------------------%
%------------------------------------------------------------------------------------------------------------------------------------------------------------------------------------------------------------------------------------------------%

%%%%%%%%%%%%%%%%%%%%%%%%%%%%%%%%%%%%%%%%%%%%%%%%%%%%%%%%%%%%%%%%%%%%%%%%%%%%%%%%%%%%%%%%%%%%%%%%%%%%
%\subsection{Exploring the range of parameters}\label{subsec:sixth}
\subsection{Exploring the interplay between opacity, base luminosity, and mass accretion rate}
\label{subsec:sixth}
%%%%%%%%%%%%%%%%%%%%%%%%%%%%%%%%%%%%%%%%%%%%%%%%%%%%%%%%%%%%%%%%%%%%%%%%%%%%%%%%%%%%%%%%%%%%%%%%%%%%

Having found that a strong increase in the opacity, essentially the electron scattering part, is able to quench the bursts at the right $\dot{M}$, we here explore in more details the sensitivity of this proposed mechanism to the two other basic parameters of the simulations: the base luminosity $L_b$, or equivalently $Q_b$, and the mass accretion rate $\dot{M}$.
We consider cases with the whole opacity multiplied by 2, 4, 6, and 8, and
a wide range of values of $L_b$, listed in Table \ref{tab:lb_qb_2}. For each of these we apply
three values of $\dot{M}$ covering the estimated range in which the quenching of burst is occurring: $0.2\dot{M}_\mathrm{Edd}$, $0.3\dot{M}_\mathrm{Edd}$, and $0.4\dot{M}_\mathrm{Edd}$.

\begin{figure*}
% FIGURE 6
\includegraphics[width=0.99\textwidth]{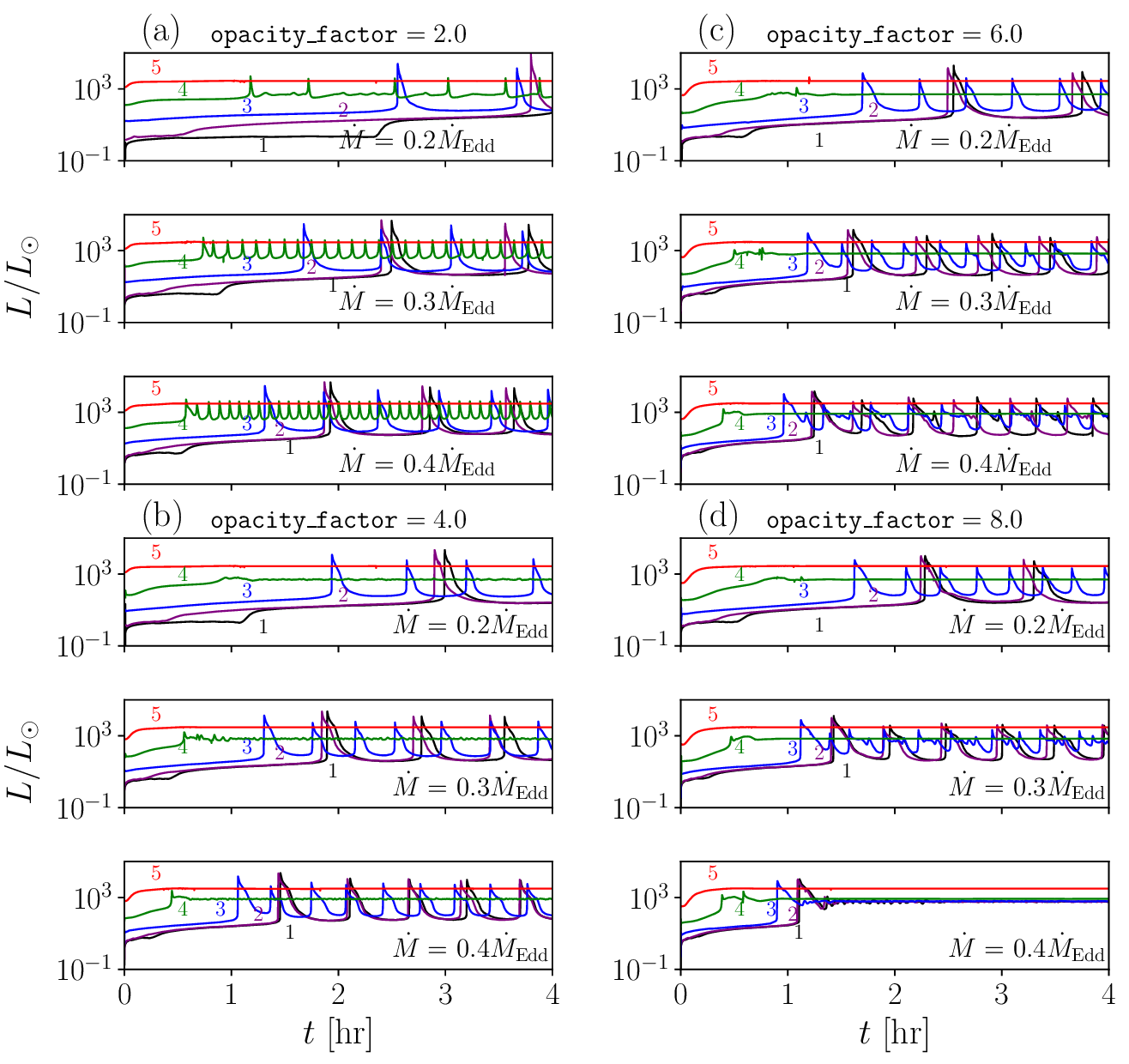}
\caption{Bursting and quenching sensitivity to opacity at various mass accretion rates and base heat flows $L_b$.
The whole opacity is scaled by a factor 2 in panel (a), 4 in panel (b), 6 in panel (c) and 8 in panel (d).
The three frames in each panel correspond to different mass accretion rates as indicated.
Curves labeled 1 to 5 correspond to different values of $L_b$ as listed in
Table \ref{tab:lb_qb_2}.
}
\label{fig:sevparams1}
\label{FIG:SEVPARAMS1}% double label to overcame the style wrong implementation in table that makes everything capital, and then a ref in a table cannot find this label
\end{figure*}

Results of these many combinations are displayed in Fig. \ref{fig:sevparams1}.
At the highest level of $L_b$ no burst ever appears because the envelope is way too hot.
At the second highest (model 4 in Table \ref{tab:lb_qb_2}), we see that with an opacity factor of 2, panel (a), the frequency of bursts sharply increases with $\dot{M}$, and millihertz oscillations are clearly displayed at $0.4\dot{M}_\mathrm{Edd}$: these have been identified as a signal of the transition from unstable to stable burning \citep{2007ApJ...665.1311H}.
When pushing the opacity factor to 4 we effectively find that burst have been completely quenched.

Concerning the three lowest cases of $L_b$, one sees that model 1 and 2 are very similar to each other at every panel regardless of $\dot{M}$ and \texttt{opacity\_factor}. 
Model 3, on the other hand, exhibits variations in the number of bursts according to the accretion rate and opacity factor. 
However, once the \texttt{opacity\_factor} is equal to 8, all three lowest levels reach a state of stable burning at $0.4\dot{M}_{\text{Edd}}$.

%------------------------------------------------------------------------------------------------------------------------------------------------------------------------------------------------------------------------------------------------%
\begin{table*}
	\centering
	\caption{Base luminosity levels for the five models in Fig. \protect\ref{fig:sevparams1}, expressed in different units.}
	\begin{tabular}{|p{1.25cm}|p{2.0cm}|p{1.5cm}|p{1.5cm}|p{1.5cm}|p{1.5cm}|}
		\hline
		\multicolumn{3}{|c|}{} & \multicolumn{3}{|c|}{ $Q_{b}\ [\mathrm{MeV\ baryon}^{-1}]$}\\
		\hline
        Model \# & $\log_{10}T_{\mathrm{eff},b}\ [\mathrm{K}]$ & $L_{b}\ [L_{\odot}]$ & $0.2\dot{M}_{\mathrm{Edd}}$ & $0.3\dot{M}_{\mathrm{Edd}}$ & $0.4\dot{M}_{\mathrm{Edd}}$\\
		\hline
        1 & 5.00 & $2.5\times10^{-5}$ & $4.5\times 10^{-7}$ & $3.0\times 10^{-7}$ & $2.2\times 10^{-7}$ \\
        2 & 6.00 & $2.5\times10^{-1}$ & $4.5\times 10^{-3}$ & $3.0\times 10^{-3}$ & $2.2\times 10^{-3}$ \\
        3 & 6.50 & $2.5\times10^{1}$ & $0.450$              & $0.300$             & $0.220$             \\
        4 & 6.75 & $2.5\times10^{2}$ & $4.5$                & $3.0$               & $2.2$               \\
        5 & 7.00 & $2.5\times10^{3}$ & $45$                 & $30$                & $22$                \\
        \hline
	\end{tabular}
        \label{tab:lb_qb_2}
\end{table*}
%------------------------------------------------------------------------------------------------------------------------------------------------------------------------------------------------------------------------------------------------%

%------------------------------------------------------------------------------------------------------------------------------------------------------------------------------------------------------------------------------------------------%
%------------------------------------------------------------------------------------------------------------------------------------------------------------------------------------------------------------------------------------------------%

%%%%%%%%%%%%%%%%%%%%%%%%%%%%%%%%%%%%%%%%%%%%%%%%%%%%%%%%%%%%%%%%%%%%%%%%%%%%%%%%%%%%%%%%%%%%%%%%%%%%
\section{Discussion and Conclusions}
%%%%%%%%%%%%%%%%%%%%%%%%%%%%%%%%%%%%%%%%%%%%%%%%%%%%%%%%%%%%%%%%%%%%%%%%%%%%%%%%%%%%%%%%%%%%%%%%%%%%

In this short letter we have explored the possibility of bursting suppression, between 0.3 and 0.4 $\dot{M}_{\rm{Edd}}$, as a consequence of changing the opacity of the envelope. We first used a global \texttt{opacity}\_\texttt{factor} to change the opacity across the whole envelope profile, finding that increasing the opacity by a factor of 10 does stabilize the nuclear burning above 0.35 $\dot{M}_{\rm{Edd}}$.

We then explored which of the main three components (free-free and scattering contribution to the radiative opacity or the conductive opacity) was responsible for the bursting suppression. We found that it is essentially the electron scattering part that is able to quench the bursts, while an increment in the free-free component or the conductive part did not induce a noticeable suppression.

Burst suppression is also controlled by the assumed base luminosity $L_b = Q_b \cdot \dot{M}/m_u$ (see Table \ref{tab:lb_qb_2}) flowing into the envelope from the stellar interior.
Extremely high values of $L_b$/$Q_b$, as in model 5, totally suppress bursts
independently of the applied changes in the opacity. (Such high $Q_b$ are, though, in the range of shallow heating inferred in the case of the system MAXI J0556-332, 
\citealt{2015ApJ...809L..31D,2017ApJ...851L..28P,2022ApJ...933..216P}, even if most of this heat actually flows toward the interior and not into the envelope.)
In the other extreme cases of vanishingly small values of $L_b$/$Q_b$, as models 1 and 2, an opacity enhancement of 8 is needed to induce burst suppression above 0.4 $\dot{M}_{\rm{Edd}}$ and intermediate cases are shown in Fig. \ref{fig:sevparams1}.

While our numerical simulations need an increased opacity for suppressing the bursts it is unlikely that electron scattering in itself could be increased by such a large factor. However, this could be seen only as a proxy replacement for an actual physical process enhancing the opacity of the envelope. This process should be acting at densities below $\sim 10^{5}$ g cm$^{-3}$, because we found that changing the electron scattering part of the opacity is sufficient and this is the density range where that is the dominating process (see Fig. \ref{fig:kappa6}).

On the other hand, a positive aspect of our results, considering that the rp-process produces nuclei with charge much above the values calculated in the Los Alamos tables \citep{2016ApJ...817..116C}, is that even large uncertainties on the free-free contribution to the  opacity at the high densities where it dominates have only very little effect on the bursting behavior predictions.

%%%%%%%%%%%%%%%%%%%%%%%%%%%%%%%%%%%%%%%%%%%%%%%%%%%%%%%%%%%%%%%%%%%%%%%%%%%%%%%%%%%%%%%%%%%%%%%%%%%%

%%%%%%%%%%%%%%%%%%%% REFERENCES %%%%%%%%%%%%%%%%%%

\bibliography{ms.bib} % if your bibtex file is called example.bib

%%%%%%%%%%%%%%%%%%%%%%%%%%%%%%%%%%%%%%%%%%%%%%%%%%%%%%%%%%%%%%%%%%%%%%%%%%%%%%%%%%%%%%%%%%%%%%%%%%%%%%%%%%%%%%%%%%%%%%%%%%%%%%%%%%%%%%%%%%%%%%%%%%%%%%%%%%%%%%%%%%%%%%%%%%%%%%%%%%%%%%%%%%%%%%%%%%%%%%%%
\appendix
%%%%%%%%%%%%%%%%%%%%%%%%%%%%%%%%%%%%%%%%%%%%%%%%%%%%%%%%%%%%%%%%%%%%%%%%%%%%%%%%%%%%%%%%%%%%%%%%%%%%%%%%%%%%%%%%%%%%%%%%%%%%%%%%%%%%%%%%%%%%%%%%%%%%%%%%%%%%%%%%%%%%%%%%%%%%%%%%%%%%%%%%%%%%%%%%%%%%%%%%

%%%%%%%%%%%%%%%%%%%%%%%%%%%%%%%%%%%%%%%%%%%%%%%%%%%%%%%%%%%%%%%%%%%%%%%%%%%%%%%%%%%%%%%%%%%%%%%%%%%%
\section{Impact of several parameters}\label{sec:AppA}
%%%%%%%%%%%%%%%%%%%%%%%%%%%%%%%%%%%%%%%%%%%%%%%%%%%%%%%%%%%%%%%%%%%%%%%%%%%%%%%%%%%%%%%%%%%%%%%%%%%%

%------------------------------------------------------------------------------------------------------------------------------------------------------------------------------------------------------------------------------------------------%
\begin{figure}
\includegraphics[width=0.40\textwidth]
    {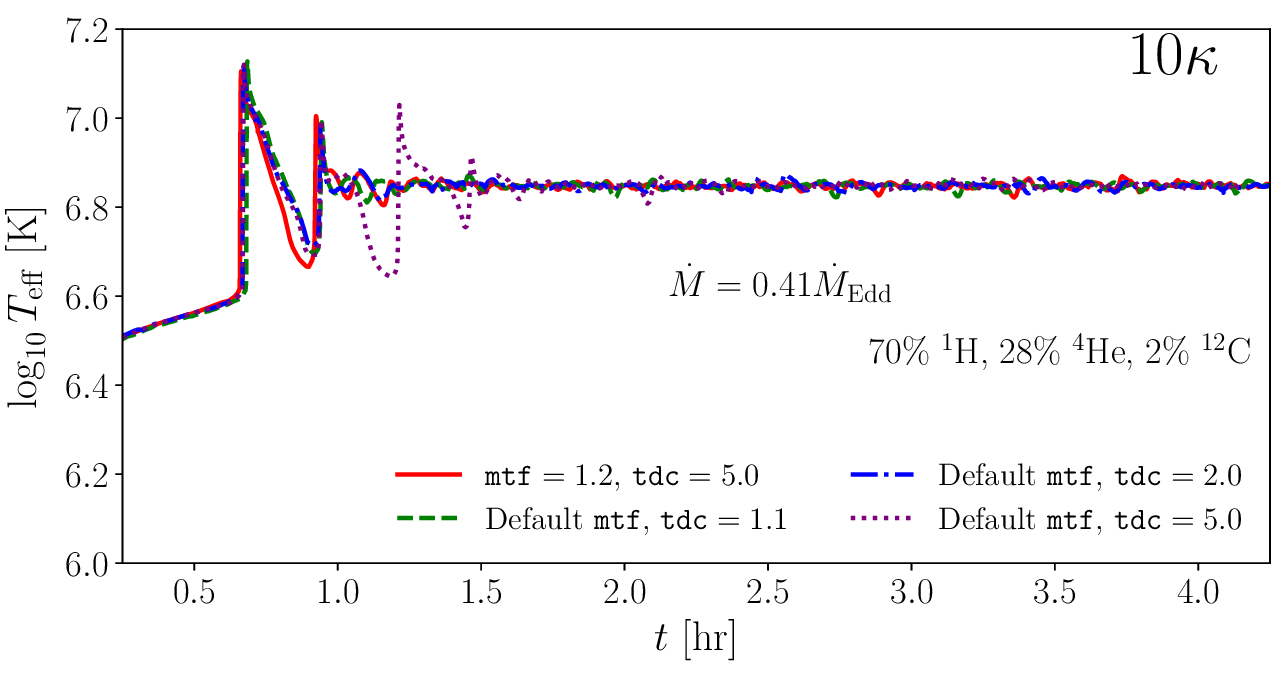}
 \caption{Effective temperature as a function of time, at fixed opacity factor and accretion rate, for different setups of timestep. 
 Here we use $\log_{10}\rho_{b} = 9.57$, $g_{s,14} = 2.16$, \texttt{mesh} $= 5.0$, $^{80}$Kr at the base, \texttt{approx140} and the composition of the accreted matter is indicated in the plot.}
 \label{fig:kappa3}
\end{figure}
%------------------------------------------------------------------------------------------------------------------------------------------------------------------------------------------------------------------------------------------------%

\begin{figure}
\includegraphics[width=0.40\textwidth]
    {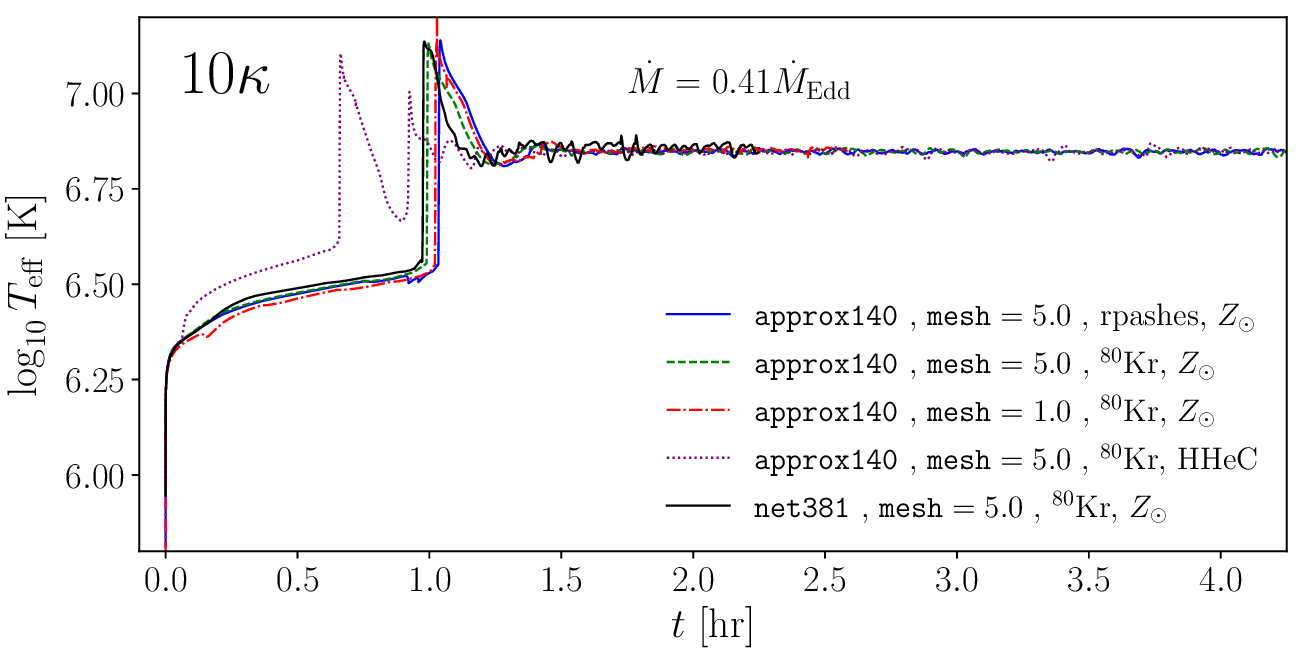}
 \caption{Effective temperature as a function of time, at fixed opacity factor and accretion rate, for different combinations of parameters. For each model the network, \texttt{mesh} parameter, composition at the base and of accreted material is indicated in the labels, HHeC corresponding to the same distribution of H, He and C as in Fig.~\ref{fig:kappa3}. Common to all models are: $g_{s,14} = 2.16$, $\log_{10}\rho_{b} = 9.57$.}
 \label{fig:kappa4}
\end{figure}

One concern could be that the bursting suppression is also a consequence of our other choices for the parameters, not only of the change in opacity. For example, setting \texttt{mtf} $= 1.2$ leads to progressively longer time steps, which may influence the evolution of the column by skipping important variations in the burning. To test whether this is the case, we ran a few simulations, keeping the accretion rate fixed at $7.26 \times 10^{-9}\ M_{\odot}\, \rm{yr}^{-1}$ ($\sim 0.4\dot{M}_{\rm{Edd}}$) as well as an increased opacity of $10\kappa$, but changing other parameters. 
In particular, we focused on the following, and the resulting light curves can be found in Figs.~\ref{fig:kappa3} and \ref{fig:kappa4}:
\begin{itemize}
    \item \textbf{Limits to the time step in the mesh.} We modified both \texttt{time}\_\texttt{delta}\_\texttt{coeff} and \texttt{min}\_\texttt{timestep}\_\texttt{factor} to favour longer or shorter time steps. 
    \item \textbf{Number of cells in the mesh.} This is partially controlled by the user via the \texttt{mesh}\_\texttt{delta}\_\texttt{coeff} parameter. In our simulations, a value around and above 5 restricts the amount of cells below 500, while with a value around 1 the cells can be as many as 2000 to 3000.
    \item \textbf{Number of species in the network.} To rule out the possibility of the electroweak reactions from the omitted nuclides in the \texttt{approx140} network altering the bursting behavior, we employed the network of 380 species described in \citet{2024arXiv240313994N}, adding neutrons for a fair comparison with the \texttt{approx140}, and thus resulting in 381 species. We refer to this larger network as \texttt{net381}.
    \item \textbf{Composition of the base.} We chose two compositions: either the rp-ashes mixture, or a single-species with a heavy nucleus, $^{80}$Kr, common to both networks we considered.
    \item \textbf{Composition of accreted material.} We explored two compositions: the first is the Solar-like distribution, detailed in Section 2, while the second composition is a simpler mixture of $70\%$ $^{1}$H, $28\%$ $^{4}$He and $2\%$ $^{12}$C, key species in the actual synthesis of heavier elements via the rp-process. $^{12}$C is necessary for a fair comparison between \texttt{net381} and \texttt{approx140} due to the absence of Li, Be and B isotopes directly connecting $^{4}$He with C, N and O in the latter network.
\end{itemize}

To test the impact of the time-step controls over the stabilization, we ran simulations altering both \texttt{mtf} and \texttt{tdc}, employing the \texttt{approx140} network, $^{80}$Kr as the composition of the base and the simpler accretion mixture. 
The models discussed in the main text have \texttt{mtf} = 1.2 and \texttt{tdc} = 5.0 (this is the red curve in the Figures of this Appendix), while we test combinations with \mesa{'}s default value of \texttt{mtf} $= 0.8$ and \texttt{tdc} $= 1.1, 2, 5$. 
The overall similarity of all results with respect to our fiducial strongly suggest that our previous results are solely due the the change in opacity and not an artifact of the numerical integration. Some discrepancies can still be seen, though they don't change the conclusions. For instance, employing the default value for \texttt{mtf} in combination with a large \texttt{tdc} induces an additional burst and delays the stabilization for 0.5 hr. The equilibrium temperature, however, is similar to the rest of the models. The second difference is the time span of the decay phase after the burst peak, which is slightly shorter in the \texttt{mtf} $= 1.2$ case than in the rest of the simulations.

%------------------------------------------------------------------------------------------------------------------------------------------------------------------------------------------------------------------------------------------------%
%------------------------------------------------------------------------------------------------------------------------------------------------------------------------------------------------------------------------------------------------%

The other important numerical parameter is resolution. The difference in size of the mesh does not produce appreciable deviations, neither in the first ``numerical'' burst nor in the equilibrium temperature, suggesting that our fiducial parameters for the resolution are high enough (Fig. \ref{fig:kappa4}).

We now turn to more physical parameters (Fig. \ref{fig:kappa4}). Regarding the size of the network, although the decay after the peak of the first burst using \texttt{net381} is faster than for the \texttt{approx140} model, the overall behavior is the same: only one burst occurs, followed by a damping process finally converging to a stable state. The oscillations from the \texttt{net381} network around the equilibrium value are slightly more visible than those generated by the \texttt{approx140} network, but this difference is minor.

With respect to the base composition we observe little difference in the stabilization properties when using either of the ashes mixtures or pure $^{80}$Kr , although some differences in the rise and decay phases of the first burst are visible. The overall simulation reaches stabilization nonetheless and there is little difference in the equilibrium temperature
%\todo{luminosity - remember to redo all plots in luminosity} 
with respect to the rest of the models. When considering different accreted composition, we notice that the absence of $Z > 6$ species in the fuel material delays the stabilization process after a secondary less-energetic burst have occurred.
However, no major changes are noticeable when the burning turns stable.

%%%%%%%%%%%%%%%%%%%%%%%%%%%%%%%%%%%%%%%%%%%%%%%%%%%%%%%%%%%%%%%%%%%%%%%%%%%%%%%%%%%%%%%%%%%%%%%%%%%%
\section{Motivation of the \texttt{approx140} network}
\label{sec:AppB}
%%%%%%%%%%%%%%%%%%%%%%%%%%%%%%%%%%%%%%%%%%%%%%%%%%%%%%%%%%%%%%%%%%%%%%%%%%%%%%%%%%%%%%%%%%%%%%%%%%%%

%------------------------------------------------------------------------------------------------------------------------------------------------------------------------------------------------------------------------------------------------%
\begin{figure*}
\includegraphics[width=0.99\textwidth]
    {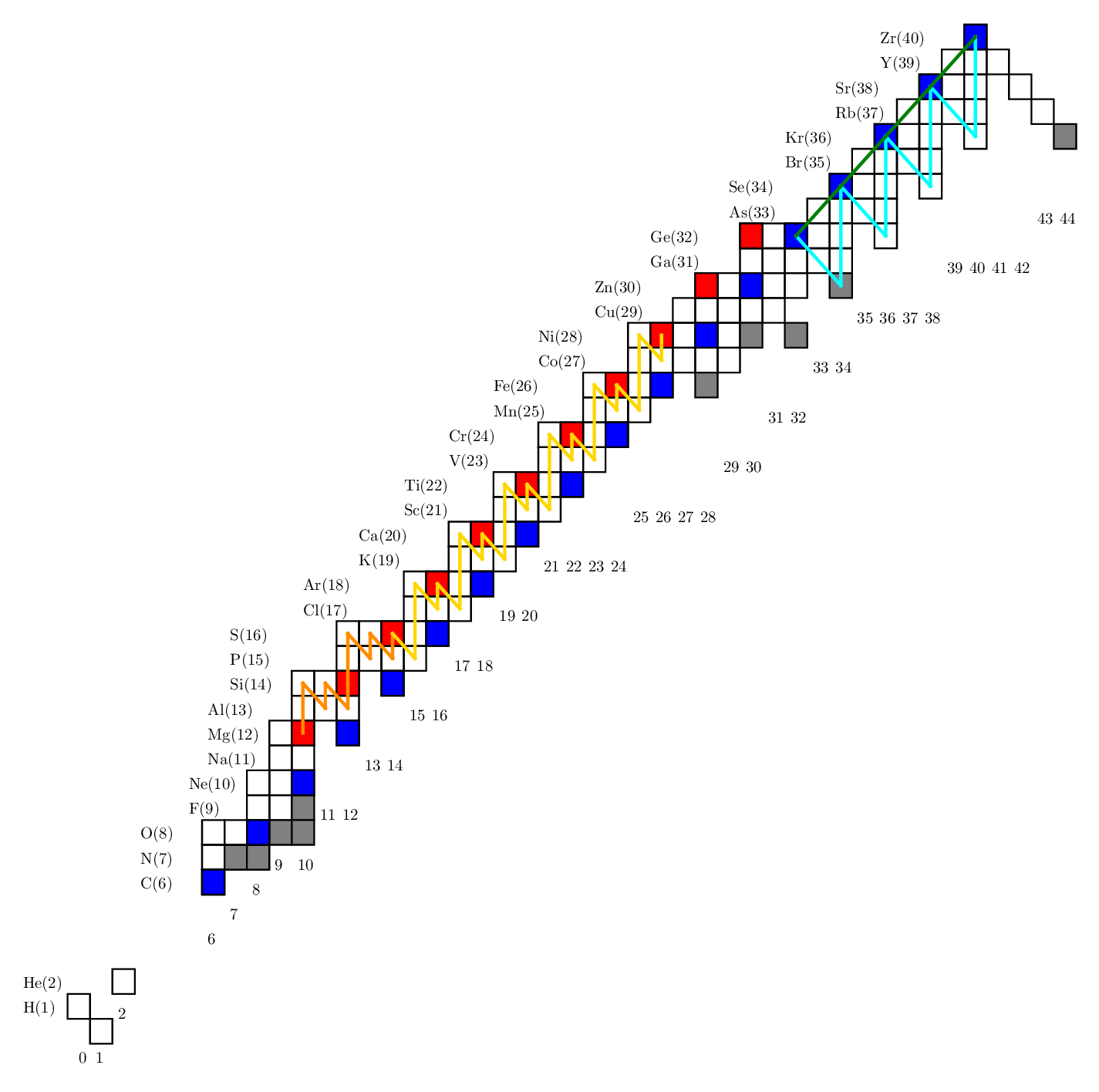}
 \caption{ Nuclide chart of the \texttt{approx140} network. Numbers in parenthesis after element symbols are the charges $Z$, while numbers below the chart are the neutron numbers 
 $N=A-Z$ of the various isotopes.
 Red squares: $T_{z} = -1$ nuclides. Blue squares: $\alpha$ nuclides. Dark-gray squares: nuclides in the valley of stability. The orange and yellow lines correspond to the main flow of the rp-process below $A = 56$, while the green and cyan lines correspond to the main flow above $A = 64$. See the main text for further details.}
 \label{fig:net140}
\end{figure*}
%------------------------------------------------------------------------------------------------------------------------------------------------------------------------------------------------------------------------------------------------%

The nuclides included in this network were listed in Table \ref{tab:net}
and the resulting flow is pictured in Fig. \ref{fig:net140}.

\begin{figure}
\includegraphics[width=0.40\textwidth]
    {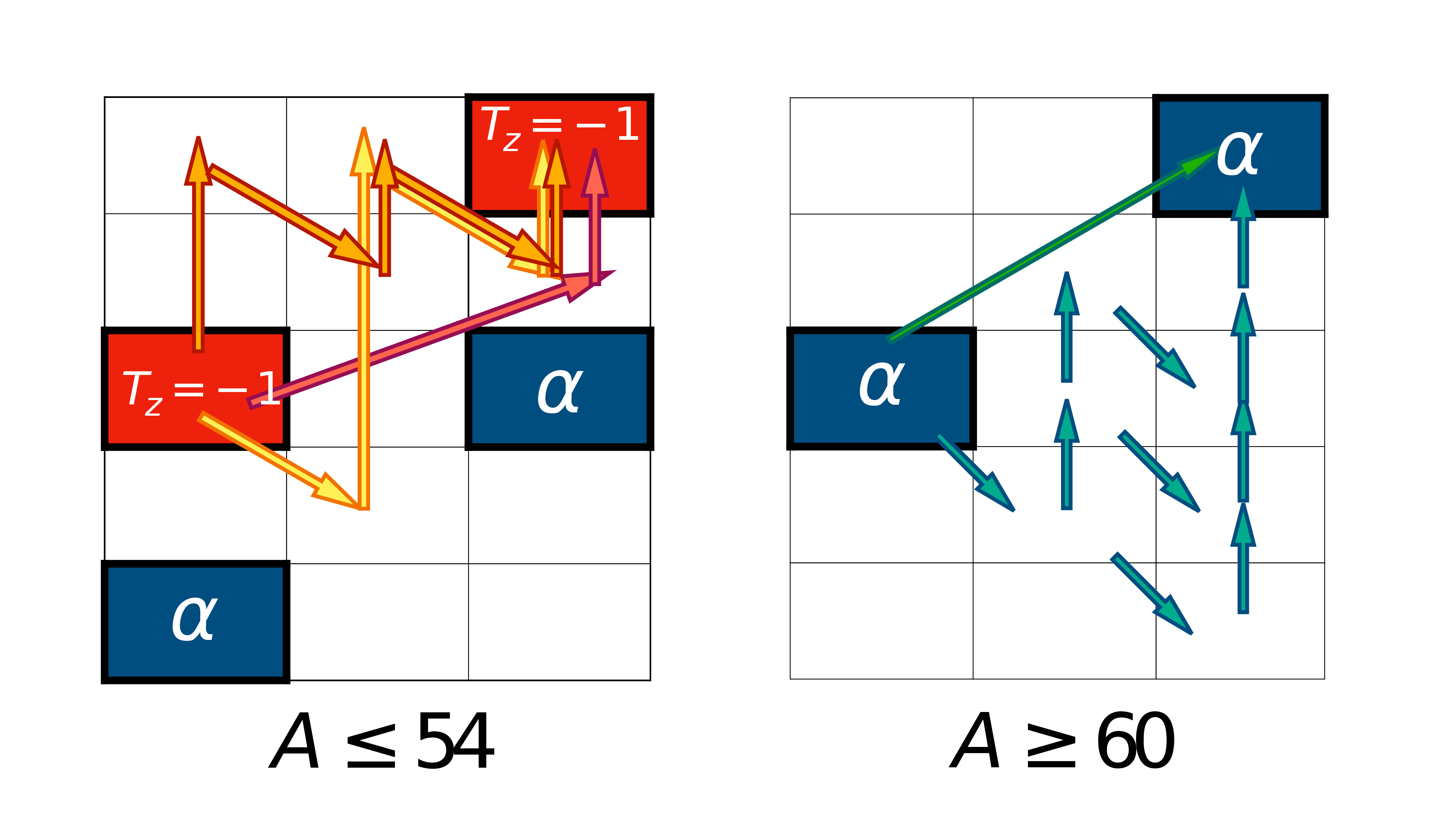}
 \caption{Left panel: box scheme of the rp-process occurring for nuclides with $A\leq 54$. Here we observe three paths of reactions, illustrated with arrows: $(a,p)-(p,\gamma)$ in red, sawtooth in orange and $\beta$-3p-$\beta$-p in yellow. Right panel: triangle-like structure of the rp-process for $A\geq 60$ nuclides. The green arrow illustrates the $(\alpha,\gamma)$ capture while the cyan arrows illustrate the paths of proton captures and $\beta^+$ decays.}
 \label{fig:triangle_box}
\end{figure}

As a first step, pp chains were omitted since the energy production occurs just at the very surface, where compression is actually more energetic than nuclear reactions. Li, Be and B nuclides are thus ignored. 
All C, N and O isotopes for the hot CNO cycle are included. For $A \leq 54$, $\beta^{+}$ decays take less than 1 hour to occur (a remarkable exception is $^{26}$Al). 
This implies that once a proton-rich $A$ isotope has been synthesized during the rp-process, it might decay towards the valley of stability in just a few minutes. 
Due to the local maxima at $\alpha$-nuclides in the distribution of ashes, we make the approximation of allowing the whole chain of $\beta^{+}$ decays towards the valley of stability only to isotopes leading to $\alpha$-nuclides. 
Considering that the integrated flow of the rp-process, either in stable or explosive burning, proceeds between isospin $T_{z} \equiv (2Z-A)/2 =-1$ nuclides and $\alpha$-nuclides, such approximation should not affect the evolution of ashes well below $10^{8}$ g cm$^{-3}$. 
A secondary consequence of the necessity of keeping $\alpha$-nuclides in the network is to adequately simulate $^{4}$He burning. Between $^{22}$Mg and $^{54}$Ni we have a ``box scheme'' \citep{1997ApJ...484..412R}, where a competition between the three chains of reactions pointed out by \citet{Fisker_2006}, occurring between $T_{z}=-1$ isotopes, takes place (see Fig. \ref{fig:triangle_box}, left). 
To decide which chain is more relevant, we considered the results of 
\citet{1999ApJ...524.1014S,2001PhRvL..86.3471S} 
as well as our own simulations (with updated reaction rates from
\citealt{Cyburt_2010} and \citealt{jinareaclib}): from $^{22}$Mg to $^{26}$Si and to $^{30}$S, the saw-tooth path is the most prominent one and thus the associated isotopes are included in the network. On the other hand, from $^{30}$S to $^{54}$Ni, the $\beta$-3p-$\beta$-p path is the dominant one. 

Between the $T_{z}=-1$ nuclides $^{54}$Ni and $^{62}$Ge, as well as the $\alpha$ nuclide $^{60}$Zn, the fictitious axis of the main flow moves from $T_{z}=-1$ to $\alpha$ nuclides and those at two $\beta^{+}$ decays of separation from them, as for instance $^{60}$Ni and $^{64}$Zn. 

At $A=64$, specifically at $^{64}$Ge synthesized during the peak of the bursts, the main flow now follows a triangular-like structure (a cascade of proton captures and $\beta^{+}$ decays connecting $\alpha$-nuclides, Fig. ~\ref{fig:triangle_box} right), while heavy isotopes such as $^{64}$Zn, $^{68}$Ge and $^{72}$Se are synthesized and do not further decay due to their long lifetimes. 

To reduce as much species in the network as possible, we simulate an endpoint to the rp process following two basic criteria: (i) H is fully exhausted at around $10^{6}$ g cm$^{-3}$, and (ii) the heaviest nuclide must have a large (above 6 days) lifetime against $\beta^{+}$ decay. We find $^{80}$Kr, and specifically the $A=80$ family, as a suitable artificial endpoint for the process.

%------------------------------------------------------------------------------------------------------------------------------------------------------------------------------------------------------------------------------------------------%
%------------------------------------------------------------------------------------------------------------------------------------------------------------------------------------------------------------------------------------------------%

%%%%%%%%%%%%%%%%%%%%%%%%%%%%%%%%%%%%%%%%%%%%%%%%%%%%%%%%%%%%%%%%%%%%%%%%%%%%%%%%%%%%%%%
%%%%%%%%%%%%%%%%%%%%%%%%%%%%%%%%%%%%%%%%%%%%%%%%%%%%%%%%%%%%%%%%%%%%%%%%%%%%%%%%%%%%%%%
%%%%%%%%%%%%%%%%%%%%%%%%%%%%%%%%%%%%%%%%%%%%%%%%%%%%%%%%%%%%%%%%%%%%%%%%%%%%%%%%%%%%%%%
%%%%%%%%%%%%%%%%%%%%%%%%%%%%%%%%%%%%%%%%%%%%%%%%%%%%%%%%%%%%%%%%%%%%%%%%%%%%%%%%%%%%%%%
%%%%%%%%%%%%%%%%%%%%%%%%%%%%%%%%%%%%%%%%%%%%%%%%%%%%%%%%%%%%%%%%%%%%%%%%%%%%%%%%%%%%%%%
%%%%%%%%%%%%%%%%%%%%%%%%%%%%%%%%%%%%%%%%%%%%%%%%%%%%%%%%%%%%%%%%%%%%%%%%%%%%%%%%%%%%%%%
%%%%%%%%%%%%%%%%%%%%%%%%%%%%%%%%%%%%%%%%%%%%%%%%%%%%%%%%%%%%%%%%%%%%%%%%%%%%%%%%%%%%%%%
%%%%%%%%%%%%%%%%%%%%%%%%%%%%%%%%%%%%%%%%%%%%%%%%%%%%%%%%%%%%%%%%%%%%%%%%%%%%%%%%%%%%%%%
%%%%%%%%%%%%%%%%%%%%%%%%%%%%%%%%%%%%%%%%%%%%%%%%%%%%%%%%%%%%%%%%%%%%%%%%%%%%%%%%%%%%%%%
%%%%%%%%%%%%%%%%%%%%%%%%%%%%%%%%%%%%%%%%%%%%%%%%%%%%%%%%%%%%%%%%%%%%%%%%%%%%%%%%%%%%%%%
%%%%%%%%%%%%%%%%%%%%%%%%%%%%%%%%%%%%%%%%%%%%%%%%%%%%%%%%%%%%%%%%%%%%%%%%%%%%%%%%%%%%%%%

\end{document}